# Machine learning accelerated finite-field simulations for electrochemical interfaces

*Chaoqiang Feng[1] and Bin Jiang[2,3]\**

[1]Hefei National Research Center for Physical Sciences at the Microscale, University of Science and Technology of China, Hefei, Anhui 230026, China

[2]State Key Laboratory of Precision and Intelligent Chemistry, Department of Chemical Physics, University of Science and Technology of China, Hefei, Anhui 230026, China

[3]Hefei National Laboratory, University of Science and Technology of China, Hefei, 230088, China.

\*Corresponding author: bjiangch@ustc.edu.cn

**Abstract**

Electrochemical interfaces are of fundamental importance in electrocatalysis, batteries, and metal corrosion. Finite-field methods are one of most reliable approaches for modeling electrochemical interfaces in complete cells under realistic constant-potential conditions. However, previous finite-field studies have been limited to either expensive ab initio molecular dynamics or less accurate classical descriptions of electrodes and electrolytes. To overcome these limitations, we present a machine learning-based finite-field approach that combines two neural network models: one predicts atomic forces under applied electric fields, while the other describes the corresponding charge response. Both models are trained entirely on first-principles data without employing any classical approximations. As a proof-of-concept demonstration in a prototypical Au(100)/NaCl(aq) system, this approach not only dramatically accelerates fully first-principles finite-field simulations but also successfully extrapolates to cell potentials beyond the training range while accurately predicting key electrochemical properties. Interestingly, we reveal a turnover of both density and orientation distributions of interfacial water molecules at the anode, arising from competing interactions between the positively charged anode and adsorbed $Cl^-$ ions with water molecules as the applied potential increases. This novel computational scheme shows great promise in efficient first-principles modelling of large-scale electrochemical interfaces under potential control.

# 1. INTRODUCTION

Understanding electrochemical interfaces at the microscopic level is crucial for elucidating fundamental electrochemical processes in electrocatalysis, batteries, and corrosion. When the electrode potential deviates from the potential of zero charge (PZC), excess charge accumulates on the electrode surface and is counterbalanced by an equal and opposite ionic charge in the electrolyte, leading to the formation of an electric double layer (EDL). The structure of the EDL, including the distribution and orientation of ions and solvent molecules, determines the electrostatic potential (ESP) profile and the electric field at the interface, thereby significantly influencing the energetics and kinetics of electrochemical reactions. Various electrochemical characterization techniques have allowed in-situ investigations of the species and their compositions within EDLs, which however cannot provide more detailed atomistic structural information. As a result, theoretical, particularly first-principles models are essential to obtain atomic-level insights into EDLs[1-6].

A critical issue in molecular modeling of electrified interfaces is how to control the electrode potential, or equivalently, how to model a voltage across two electrodes[7-8]. In general, three different categories of first-principles approaches have been developed: grand-canonical[9-12], counter-ions[3, 13-15] and finite-field methods[16]. The grand-canonical approach directly controls the Fermi energy level and electrode potential by varying the number of electrons, but it requires an implicit solvent model to maintain electroneutrality in the EDL[17]. The counter-ions method charges the electrode by introducing hydrogen, alkali metal, or halogen atoms near the interface, but it in principle controls the surface charge density only rather than the electrode potential during simulations[7]. Moreover, it is technically challenging to achieve a continuous variation in interfacial charge density because the number of ions can only be changed discretely, making it difficult to precisely tune the electrode potential to desired values[15]. In contrast, finite-field methods, including both constant electric field (**E**)[18-19] and constant electric displacement (**D**)[20] formulations, have been successfully applied to a wide range of electrode/electrolyte systems[16, 21-24]. A major advantage of finite-field methods is ensuring simulations of both electrode interfaces in a full cell at constant potentials. In addition, the charge response at the electrode surface arises from the polarization of free electrons within the metal electrode, naturally satisfying electroneutrality.

Applying finite-field methods to model the EDL of a realistic heterogeneous system needs to overcome the large temporal and spatial scales[25]. For example, the ionic charge distribution within the electrolyte equilibrates in the ns scale, which exceeds the capabilities of ab initio molecular dynamics (AIMD) based finite-field simulations that typically are limited to 10~100 ps[16]. Classical molecular mechanics (MM) approaches have been employed to characterize intermolecular interactions in bulk electrolytes, however, they are less accurate in describing the electronic structure of metal electrodes and metal–electrolyte interfaces in the presence of electric fields[26-27]. An alternative strategy is to employ hybrid quantum mechanical/molecular mechanics (QM/MM) approaches, *e.g.* by performing QM calculations on the electrode subsystem only. This scheme cannot however account for the charge transfer between the electrode and the electrolyte.

Machine learning potentials (MLPs) have been increasingly used in accelerating molecular dynamics (MD) simulations with first-principles accuracy. Several MLPs for simulating electrified interfaces have been combined with the classical Siepmann-Sprik model, in which electrodes follow Gaussian charge distributions with fluctuating magnitudes, while electrolytes are represented by point charge[28]. For example, Zhang and co-workers[29-32] employed an atomistic ML approach (PiNN) to compute the charge response kernel and base charge, which were then used to generate response charges at electrode sites and to propagate classical MD simulations. Similarly, Grisafi et al.[33-34] developed the symmetry-adapted learning of three-dimensional electron densities (SALTED) approach to predict the electronic density of metal electrodes and charge response under finite-field perturbations, so as to compute electrostatic forces that drive the dynamics of the metal-electrolyte interface. In contrast, Zhu et al.[35] adopted the Deep Wannier model and classical Siepmann-Sprik model to describe the dielectric response in the electrolyte and metal electrode, respectively, allowing the estimation of long-range electrostatic energy. The remaining short-range energy was separately learned using another local atomic descriptor-based MLP model. However, these studies retained the classical description of either electrodes or electrolytes.

In this work, we leverage two ML models to accelerate fully first-principles finite-field simulations of electrified interfaces and associated electrochemical properties. This ML-based finite-field approach includes a field-dependent MLP model for the metal-electrolyte interface to enable machine learning molecular dynamics (MLMD) simulations under constant potential[36] and a ML electron density response (MLEDR) model to predict charge transfer both within the electrode and between the electrode and electrolyte at the electrochemical interface under an applied electric field[37]. As a proof-of-concept, this approach is validated in a benchmark system, i.e. the Au(100)/NaCl(aq) interface. Our ML models accurately capture not only potential-dependent interfacial structures as well as the charge response and differential capacitance. This ML scheme offers first-principles accuracy as AIMD simulations, while achieving speedups of several orders of magnitude and eliminating the need for classical description of the metal or electrolytes, making it a powerful tool for studying complex electrochemical systems.

## 2. METHODS

### 2.1. Machine Learning models for field-dependent potential energy surfaces and electron responses

To characterize the structure and dynamics of electrified interfaces under finite-field conditions, an accurate global potential energy surface (PES) capturing system-field interactions is essential. Although finite-field MD based on classical force fields[38] or density functional theory[16] (DFT) has been proposed, these methods suffer from either limited accuracy or low computational efficiency. To this end, we adopted the field-induced recursively embedded atom neural network (FIREANN) approach[36], which employs the field-induced embedded atom density (FI-EAD) descriptor to characterize both atomic environments and system-field interactions. The FI-EAD descriptor comprises two components—Gaussian-type orbitals (GTOs) that depend on neighboring atoms, and a field-dependent orbital,

$$\psi^m_{l_x l_y l_z} = \sum_{j \neq i}^{N_c} c_j \varphi^m_{l_x l_y l_z}(\vec{r}_{ij}) + c_\varepsilon \varphi_{l_x l_y l_z}(\vec{\varepsilon}_i) \tag{1}$$

where the GTO of the $j$th neighbor atom is characterized by its center ($r_m$), width ($\sigma$), and angular momentum ($l = l_x + l_y + l_z$) as,

$$\varphi^m_{l_x l_y l_z}(\vec{r}_{ij}) = (x_{ij})^{l_x}(y_{ij})^{l_y}(z_{ij})^{l_z} \exp\left[-\frac{(r_{ij} - r_m)^2}{2\sigma^2}\right] \cdot f_c(r_{ij}) \tag{2}$$

and the field-dependent orbital contains only directional terms,

$$\varphi_{l_x l_y l_z}(\vec{\varepsilon}) = (\varepsilon_x)^{l_x}(\varepsilon_y)^{l_y}(\varepsilon_z)^{l_z} \tag{3}$$

Each FI-EAD feature is formulated as the squared linear combination of all neighboring atomic orbitals and the field-dependent orbital,

$$\rho_i^n = \sum_{l=0}^{L} \sum_{\substack{l_x, l_y, l_z \\ l_x + l_y + l_z = l}} \frac{l!}{l_x! l_y! l_z!} \left[\sum_{m=1}^{N_\varphi} d_m^n \left(\sum_{j \neq i}^{N_c} c_j \varphi^m_{l_x l_y l_z}(\vec{r}_{ij}) + c_\varepsilon \varphi_{l_x l_y l_z}(\vec{\varepsilon}_i)\right)\right]^2 \tag{4}$$

By varying these hyperparameters of GTOs, a vector of FI-EAD features is formed and serves as the input for each atomic neural network (NN) to predict atomic energy. A key advantage of the FI-EAD feature is its ability to implicitly capture three-body interactions with a computational scaling of $O(N_c)$. The combination coefficient for the $j$th neighboring atom is produced by an atomic NN conditioned on the local environment of the $j$th atom, which can be used to generate new FI-EAD features via Eq. (3) to establish a message-passing like iteration,

$$c_j^t = \mathrm{NN}_j^{t-1}\left(\rho_j^{t-1}\left(c_j^{t-1}, r_j\right)\right) \tag{5}$$

This form enables the incorporation of higher-order and nonlocal interactions, which are essential for accurately capturing electrostatic interactions at electrified interface.

Another critical factor in characterizing electrified interfaces is the charge response of the metal electrode in the presence of electrolytes and an applied electric field, which can be predicted using our recently developed MLEDR model for representing the real-space electron density distribution[37]. The architecture of the MLEDR model resembles the FIREANN model for PES but includes an additional component: ghost atoms. These ghost atoms are essentially grid points of electron density distribution obtained by DFT, whose local environments are characterized by the FI-EAD descriptor over surrounding real atoms and are mapped to charge response value through an atomic NN. Specifically, the learning target of the MLEDR model is defined as the difference between the electron density of the system with and without an applied field,

$$\Delta n = n_{metal-electrolyte}(r; \vec{\varepsilon}_z) - n_{metal}(r) - n_{electrolyte}(r) \qquad (6)$$

The learned charge response can be then used to calculate the charge density at the electrode surface, and subsequently the differential capacitance.

## 2.2. Finite Field methods, Datasets and Training setup

In this work, we employ the finite-field approach introduced by Dufils et al. to set up electrochemical interfaces by imposing a voltage between two electrodes within three-dimensional periodic boundary conditions (PBC) without the requiring the introduction of a vacuum region[39]. As illustrated in Figure 1, this method applies an electric field to a single electrode maintained at constant potential and in contact with an aqueous ionic solution. The applied field induces interfacial charging, leading to the formation of two EDLs, one on each side of the electrode. This setup has been successfully applied in simulations of electrochemical interfaces, encompassing both classical and first principles MD[16, 39].

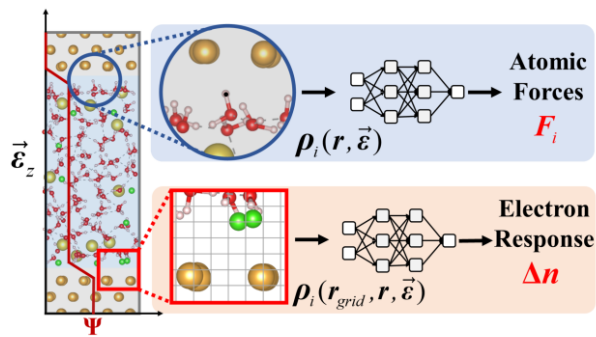

**Figure 1.** Schematic of a full cell setup to model a metal-electrolyte interface with the finite-field method, where the Au(100) electrodes are separated by an aqueous NaCl solution of 5.5 mol/L. The applied cell potential is along the $z$ axis given by $\Delta\Psi = \vec{\varepsilon}_z \cdot L_z$, with $L_z$ being the corresponding box length.

Electrochemical interfaces studied here were represented by a five-layer Au(100) periodic slab constructed with a 4×4 supercell, initially separated by a 25 Å region along the vertical direction filling with electrolytes consisting of 100 $H_2O$ molecules and 10 NaCl ions (5.5 mol/L), whose initial configurations were generated using the Packmol package[40]. The system was first equilibrated using a classical force field and constant-potential MD under applied electric fields of 0.0 V/Å, 0.029 V/Å, and 0.057 V/Å along the $z$-direction, corresponding to cell potentials of 0 V, 1 V, and 2 V, respectively. AIMD simulations under the same field strengths were then performed for 20 ps to generate the initial dataset for training. An active learning workflow was employed to progressively sample the configuration space[41]. A total of 12,139 configurations were used to train the FIREANN PES. Note that only atomic forces were used to train the PES, as energies were inconsistent due to the multivalued nature of the polarization of periodic systems under an applied field. Additionally, the charge response defined in Eq. 5 for these configurations were used to train the MLEDR model. To further demonstrate the computational efficiency of the FIREANN model, we built a larger electrochemical interface system comprising five layers Au(100) slabs with an 8×8 supercell and an electrolyte containing 800 water molecules and 80 NaCl ions, totaling 2880 atoms.

All DFT calculations were carried out using the open source CP2k/Quickstep package[42] with the Perdew-Burke-Ernzerhof (PBE) functional[43], Goedecker-Teter-Hutter (GTH) pseudopotentials[44-45], and double-ζ basis sets with one set of polarization functions (DZVP)[46]. A plane-wave energy cutoff of 800 Ry was employed, and Grimme's D3 dispersion correction was applied to capture for dispersion interactions[47]. All MLMD simulations based on the FIREANN PES were conducted with Large-scale Atomic/Molecular Massively Parallel Simulator (LAMMPS)[48] in the canonical (NVT) ensemble using a timestep of 0.5 fs. The target temperature was maintained at 330 K using a Nose–Hoover thermostat with a damping time of 500 fs. In each condition, the interfacial system was equilibrated for 1 ns to ensure relaxation of the ion distribution within the electrolyte under the applied finite field, followed by 5 ns of sampling.

## 3. RESULTS and DISCUSSION

## 3.1 Validation of MLPs for electrochemistry interface

The accuracy of the FIREANN PES is first assessed by evaluating the root mean square error (RMSE) of atomic forces predicted by FIREANN against DFT calculations under finite electric fields. Note that the total energy and the polarization of the periodic system under an external field are not well-defined targets for ML as the polarization is a multi-valued property. As listed in Table S1 of the Supporting Information (SI), the PES model achieves an RMSE for atomic forces of approximately 43 meV/Å, which is comparable to values reported in previous studies[35-36]. Additionally, the validation dataset shows similar RMSEs as the training dataset. To further assess the generalizability of the model beyond the validation dataset, we monitor the electrolyte polarization varying with the cell potential during MD simulations. As shown in Figure 2, as the cell potential varies from 0 V to 4 V, to -4 V, and back to 0 V, by 0.5 V per 0.5 ns, the total electrolyte polarization follows accordingly and sufficiently. The electrolyte polarization consists of two components — the orientational relaxation of solvent water molecules and the directional transport of solute ions. Here the latter dominates the polarization response due to the high electrolyte concentration. Importantly, although the training dataset includes data points with three discrete cell potentials only, the model accurately covers the electrolyte polarization in the entire range of [-4 V, 4 V]. This demonstrates both interpolation and extrapolation capabilities of the FIREANN PES, allowing continuous variation of the cell potential in atomistic simulations.

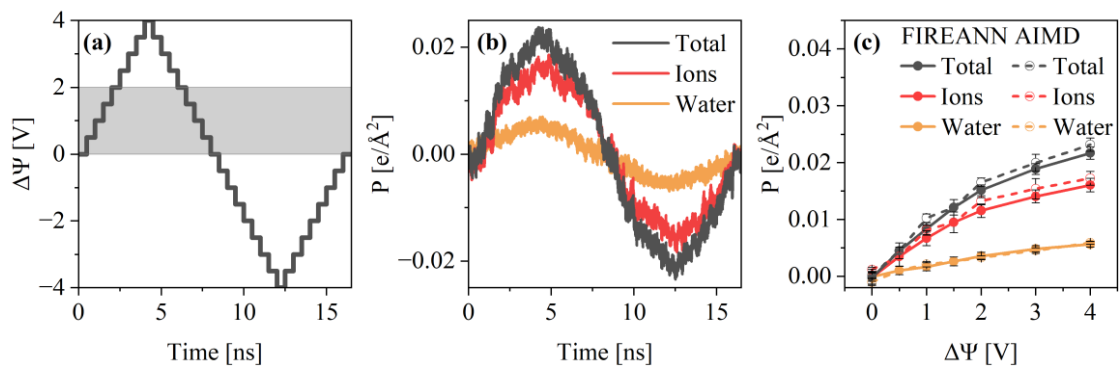

**Figure 2.** Variation of electrolyte polarization under the varying cell potential for the Au(100)/NaCl(aq) interface system. The grey area in panel (a) indicates the range of cell potentials included in the training dataset of FIREANN model. Panel (b) illustrates two contributions to electrolyte polarization: ion transfer (red) and water molecular orientation (yellow). Panel (c) compares the variation of two components of electrolyte polarizations obtained by FIREANN-based MD and AIMD simulations. FIREANN results were repeated three times to estimate the standard deviations.

Figure 2c compares the electrolyte polarizations at different potentials obtained from AIMD and from MLMD simulations. It should be noted that the typical timescale of AIMD simulations (*e.g.*, 10-100 ps) is insufficient for the electrolyte to fully relax, as shown in Figure S1 of SI, all these results were obtained by first performing equilibrated MLMD simulations for 1 ns with the FIREANN PES for a given cell potential (in place of the classical MD simulations in literature), followed by 5 ns MLMD simulations and 30 ps AIMD simulations. The total electrolyte polarization and individual contributions from water and ions with MLMD simulations agree well with AIMD results, demonstrating the accuracy of our FIREANN PES.

## 3.2 Differential Helmholtz Capacitance

The surface excess charge density on the electrode directly determines the strength of the interfacial electric field, the orientation of interfacial water molecules, and the hydration states of ions. In finite-field methods, the surface excess charge density is induced by the applied electric field and depends on the electrolyte configuration, while the average charge response correlates with the cell potential. The Helmholtz capacitance, which characterizes the surface charge response of an EDL system as a function of the cell potential, can be computed using the following formula,

$$C_H = 2 \cdot \left( \frac{d\sigma_m}{d\Delta\Psi} \right) \tag{7}$$

where $\Delta\Psi$ is the sum of the potential differences across both sides of the metal electrode, and $\sigma_m$ is the surface charge response density imposed on the metal electrode surface. A total of 2000 configurations were extracted from MLMD simulations for each cell potential, and the corresponding charge responses were predicted using the MLEDR model. Notably, charge transfer from the electrolyte to the electrode occurs at the cell potential of 0 V. Further charge analysis reveals that this transfer originates from chloride ions (Cl⁻) specific adsorbed on the electrode surface. As the applied electric field increases, charge transfer occurs within the electrode, resulting in the formation of two electrified interfaces. The interfacial charge transfer is quantified by integrating the charge response from the central layer of the metal slab toward the left or right electrolyte region. As illustrated in Figure 3b, the Helmholtz capacitance is calculated with a maximum value of approximately 20.8 $\mu F/cm^2$ at the cell potential of 0 V. The capacitance obtained in this study is lower than those reported in previous finite-field AIMD simulations[16], which may be attributed to the higher electrolyte concentration used in this work. Our result is consistent with the prediction of Guo *et al.*[49] in similar Au(100)/electrolyte systems based on an ML accelerated counter-ions approach, with comparable capacitances of 17-19 $\mu F/cm^2$. However, additional DFT single-point calculations are required for post-processing to estimate averaged work function and surface charge density in that work. It is worth noting that changing the sign of the cell potential reflects the change of applied field direction, which changes the side of the anode and cathode but would not change the overall dynamics. This feature is fully captured by our FIREANN PES. Hence, only positive cell potentials are discussed hereafter, unless otherwise stated.

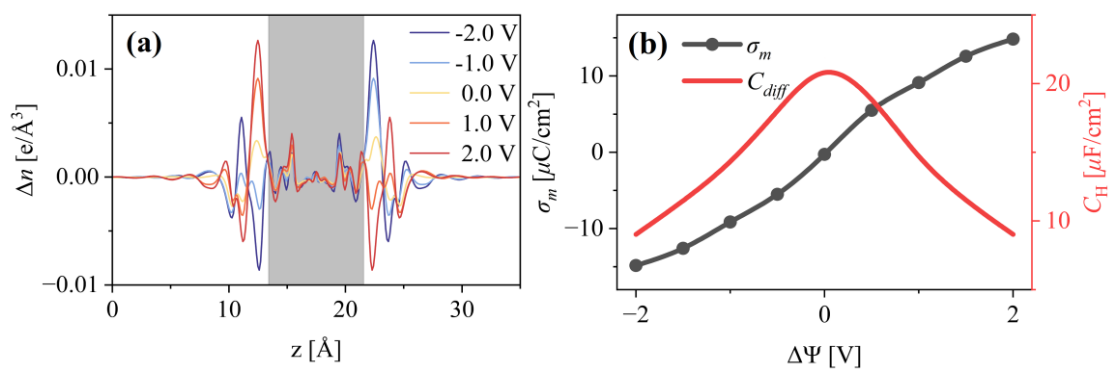

**Figure 3.** (a) Response of charge density averaged over the *xy* plane and (b) the cell potential-dependent differential capacitance of the Au(100)/NaCl(aq) interface with the salt concentration of 5.5 mol/L. The grey area in all panel indicates the Au electrode.

## 3.3 Structure of Au(100)/NaCl(aq) interface

The structure and composition of EDLs play a critical role in determining the activity and selectivity of catalytic reactions, including nitrogen reduction and oxidation[50-51], hydrogen/oxygen evolution reaction[52-54], water oxidation[55], and electro/photo chemical CO and $CO_2$ reduction reactions ($CO_{(2)}RR$)[56-58]. Atomic structures of the electrified Au(100)/NaCl(aq) interface obtained from 0 V to 4.0 V were analyzed from above MLMD trajectories during 5 ns. This extended timescale enhances the statistical reliability and enables the calculation of the smooth potential of mean force (PMF).

Figures 4a and 4b show the equilibrium ion concentration profiles varying with applied potential. As the cell potential increases, corresponding to an increase in the excess surface charge density on the electrode, oppositely charged ions accumulate near the electrode surface. This behavior is consistent with the expected ion–electrode Coulomb interactions. An interesting phenomenon is that the peak of the $Na^+$ concentration near the anodic surface hardly moves as the cell potential increases. This counterintuitive trend can be attributed to the rapid adsorption of $Cl^-$ onto the anodic surface, resulting in local charge inversion, in agreement with previous findings[16]. Furthermore, the preferential adsorption of $Cl^-$ over $Na^+$ on the electrode surface can be rationalized by the weaker solvation of larger $Cl^-$ ions by water molecules, making them more prone to direct surface adsorption. In contrast, $Na^+$ ions, being smaller and more strongly solvated, experience greater repulsive solvation forces. We note that the concentration profile of $Na^+$ exhibits multiple peaks on the cathodic surface when the cell potential is 3 or 4 V. This is likely due to the strong electrostatic interactions at high cell potentials, disrupting the $Na^+$ hydration layer, thereby allowing a fraction of Na$^+$ ions to approach the cathodic surface more closely. Additionally, the concentration distributions obtained from MLMD simulations are much less oscillatory than those from AIMD in Figure S2, owing to over 100 times longer simulation time. Indeed, AIMD with insufficient relaxation time leads to some artificial peaks of the concentration profiles, particularly in the bulk region, underscoring the advantage of MLMD simulations in analyzing atomic structures of electrode-electrolyte interfaces.

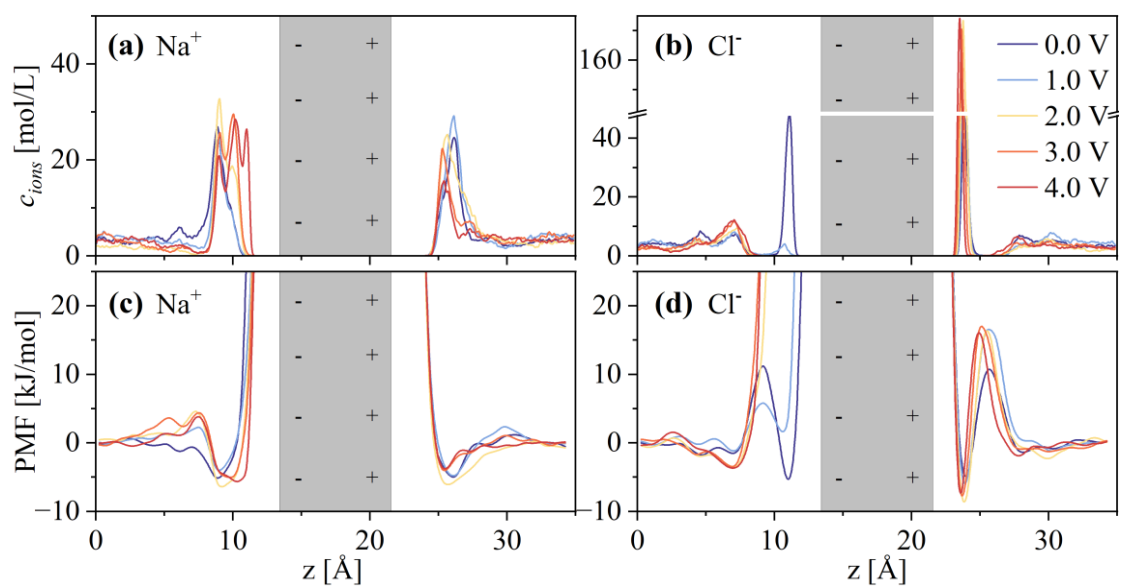

**Figure 4**. Concentration distributions and PMFs of cations (Na$^+$) and anions (Cl$^-$) as a function of the distance to the electrode under various cell potentials obtained from FIREANN simulations. The grey area in each panel represents the Au electrode being positively (anode) or negatively (cathode) charged on the respective side.

The PMFs for the approach of ions to the electrode surface is directly related to the equilibrium ion concentration profiles in Figures 4a and 4b through the following relation[59],

$$\text{PMF}(z) = -RT \ln \frac{c(z)}{c(z_0)} \tag{8}$$

where $\text{PMF}(z)$ denotes the free energy of ions at position $z$ relative to that at the bulk (defined by the midpoint of the electrolyte layer at $z_0$) along the direction normal to the interface. As shown in Figures 4c and 4d, the electrostatic repulsion drives Cl$^-$ ions away from the negatively charged electrode, while Na$^+$ ions are increasingly attracted toward it, with the increasing cell potential. A notable feature is the presence of an energy barrier for Cl$^-$ adsorption on the anode, with a potential-dependent height of approximately 10 kJ/mol. This barrier arises from the high surface coverage of Cl$^-$, which effectively screens the interaction between the positively charged electrode and Na$^+$ ions. As a result, the free energy profiles of Na$^+$ in the anodic region remain relatively unchanged across different cell potentials.

Next, we analyze the structural distributions of water molecules at the electrified Au(100)/NaCl(aq) interface. Figure 5a presents the water density distributions at different cell potentials, which consistently exhibit a pronounced peak approximately 3.1 Å from the electrode surface, indicative of a well-defined first solvation layer. Figure 5b illustrates the dipole orientation profiles, defined as $\rho_{H_2O} \cdot \cos\theta$, where $\theta$ is the angle between the bisector of $\angle HOH$ and the positive z-axis. With zero cell potential, most interfacial water molecules adapt themselves so that the O–H bond pointing toward the electrode surface. This is because of the hydration of Cl$^-$ ions that are directly adsorbed onto the electrode. As the cell potential increases, the orientation of the water dipole becomes progressively more ordered to follow the direction of the electric field.

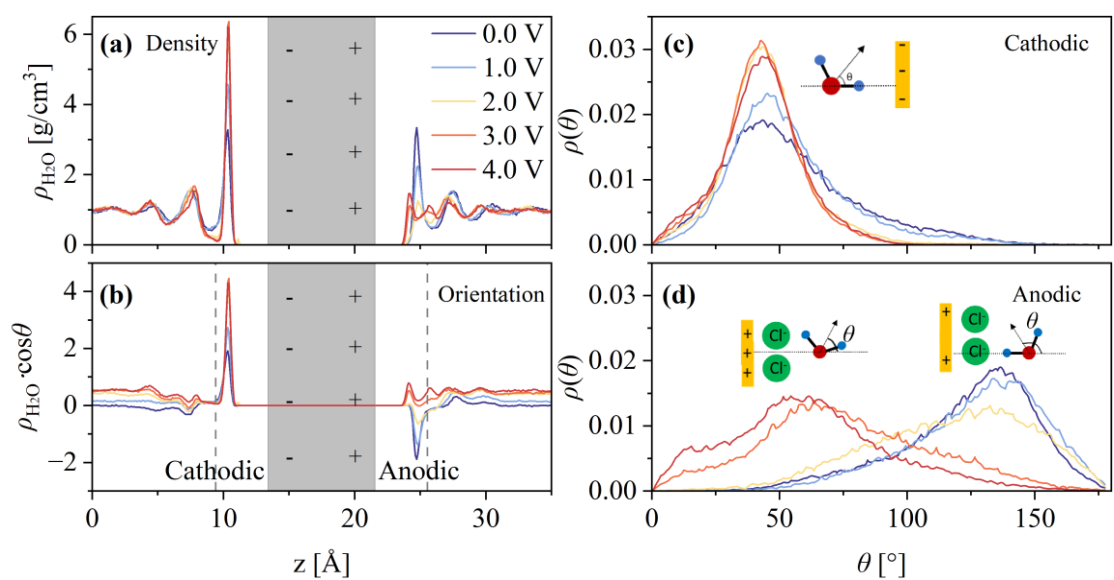

**Figure 5**. (a) Density and (b) dipole orientation distributions of water molecular along the surface normal, within the Au electrode (gray plates) cathode and anode. Normalized orientational distributions ($\theta$) for water molecules in the cathodic (c) and anodic (d) interface obtained from FIREANN simulations, where the insets show the dominant water orientation at different peaks.

More specifically, Figures 5c and 5d display the angular distributions for interfacial water molecules located within 4 Å to the cathode and anode surface, respectively. Interestingly, water molecules near the cathodic interface exhibit a pronounced peak at approximately 45°, corresponding to an orientation with one O–H bond pointing toward the electrode, as shown in the inset of Figure 5c. This configuration facilitates the formation of an extended hydrogen-bonding network with neighboring water molecules, thereby enhancing interfacial structural stability. Furthermore, this effect strengthens with increasing cell potential, leading to an increasingly sharper angular distribution. In contrast, charge inversion occurs in the anodic interface due to the excessing adsorption of Cl⁻ ions on the positively charged electrode. These anions tend to attract one of the hydrogen atoms in water molecules towards the electrode, yielding dominant angular distributions around ~135°, consistent with the previous finding[16]. At higher cell potentials (*e.g.* 3 ~ 4 V), the stronger charge response in the electrode renders more positive charge concentrated on the anodic surface, manifesting stronger interactions with interfacial water molecules, while the Cl⁻ coverage is limited by its chemical potential and cannot increase proportionally, as seen in Figure 4b and representative snapshots in Figure S2. As a result, the water molecule tends to reorient and allow its negatively charged center (*i.e.* oxygen atom) towards the anode, resulting in the significant shift of the angular distribution to smaller angles. More detailed tests show in Figure S3 that this shift starts at the cell potential of approximately 2.5 V.

Note that AIMD simulations yield similar distributions as discussed above (Figure S4), but with much higher computational costs. Moreover, the well-trained FIREANN potential allows us to study the structure and dynamics of the Au(100)/NaCl(aq) interface with a more extended cell. To this end, we performed MLMD simulations using a cell setup with 5 metallic layers and 64 atoms in each layer of the electrode, plus a thicker electrolyte layer with 50 Å, consisting of 2880 atoms. The corresponding distributions of ions and water molecules under various cell potentials are presented in Figure S5, which compare well with these obtained in the smaller cell for training.

These results clearly demonstrate the scalability of our FIREANN model to the more complex systems than trained.

## 4. CONCLUSIONS

In this work, we propose an efficient strategy to accelerate finite-field simulations electrochemical interfaces by leveraging machine learning techniques. A key feature of this strategy is using a field-induced (non-local) message-passing neural network framework to learn both atomic forces and charge responses of the metal-electrolyte system computed by density functional theory in the presence of external electric fields, thus allowing highly efficient machine learning molecular dynamics simulations and the evaluation of Helmholtz capacitance under various constant cell potentials. We validate this approach in a benchmark Au(100)/NaCl(aq) interfacial system, demonstrating its exceptional accuracy and superior computational efficiency—both in time and system size scalability—compared to conventional finite-field ab initio molecular dynamics simulations. Furthermore, this approach exhibits excellent extrapolation capabilities beyond the range of cell potentials included in the training data, allowing us to identify a turnover in the orientation distribution of interfacial water molecules at the anodic interface at high cell potentials. This reorientation is because the positively charged anode interact more strongly with water molecules than adsorbed Cl anions to as the applied potential increases. The proposed machine learning strategy offers broad potential for modelling large-scale electrochemical interfaces with first principles accuracy under potential control. It is particularly well suited for investigating metal-electrolyte interfacial structures and electrocatalytic reactions, especially when combined with enhanced sampling techniques to explore the associated free energy landscapes.

## Acknowledgement

We acknowledge the support from the Innovation Program for Quantum Science and Technology (2021ZD0303301), Strategic Priority Research Program of the Chinese Academy of Sciences (XDB0450101), and National Natural Science Foundation of China (22325304, 22221003, and 22033007). The ML simulations and model training were performed on the robotic AI-Scientist platform of Chinese Academy of Sciences.

## Supplementary Information

Additional information for the performance of the FIREANN model under different cell potential; Variation of electrolyte polarization from MLMD and AIMD; Snapshot in the finite-field MLMD trajectories under varying cell potential; Detailed results on the dipole orientation of anodic interface water under varying cell potentials; Concentration distributions of ions and density and dipole orientation distributions of water molecules from AIMD; Concentration profiles of ions, density, dipole orientation distributions of water molecules in the more extended cell.

# Code and Data availability

The original FIREANN package and the FIREANN model for the electron response module as a branch of the original package are openly available in the GitHub repository at https://github.com/bjiangch/FIREANN. Additionally, the configurations and atomic forces dataset used for the PES model are also provided. The electron response dataset used for the MLEDR model are available upon reasonable request from the corresponding author.

# Conflicts of interest

The authors declare no conflict of interest.

# Keywords

Electrochemistry; Finite-field methods; Machine learning; Molecular dynamics;

**Entry for the Table of Contents:**

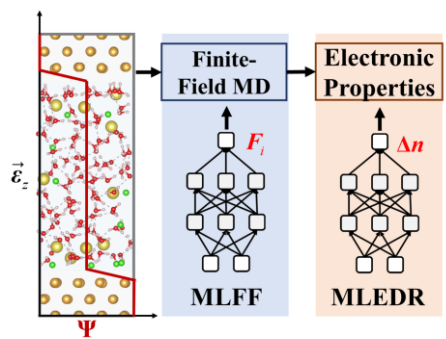

A machine learning framework to accelerate finite-field simulations of electrified interfaces, which integrates a field-dependent machine learned force field and a machine learned electronic response predictor. In a benchmark Au(100)/NaCl(aq) system, this framework shows exceptional efficiency and first-principles accuracy, as well as scalability with respect to time, system size, and electric potential in finite-filed simulations.